\documentstyle[prd,aps]{revtex}
\begin{document}
\draft
\title{THE GENERALIZED RAYCHAUDHURI EQUATIONS : EXAMPLES}
\author{Sayan Kar \thanks{ Electronic Address :
sayan@iopb.ernet.in}}
\address{Institute of Physics,\\
Sachivalaya Marg, Bhubaneswar--751005, INDIA}
\maketitle
\begin{abstract}

Specific examples of the generalized Raychaudhuri Equations
for the evolution of
deformations along families of $D$ dimensional surfaces embedded in a
background $N$ dimensional spacetime are discussed. These include
string worldsheets embedded in four dimensional
spacetimes and
two dimensional timelike hypersurfaces in a three dimensional
curved background. The issue of focussing of families of surfaces
is introduced and analysed in some detail.

\end{abstract}
\vskip 0.125 in
\parshape=1 0.75in 5.5in
PACS number(s): 11.27.+d, 11.10.Kk

\section{INTRODUCTION}

The Raychaudhuri equations for null or timelike geodesic
congruences [1] provide us with a clear picture of the evolution
of deformations along these specific families of curves.
Together with the Einstein equations and an assumption about the
nature of matter (in terms of an Energy Condition [2]) one arrives
at the Focussing theorem which states that timelike/null
geodesic congruences tend to get focussed within a finite value
of the affine parameter used in defining these curves [2,3]. The
Focussing theorem along with some other global arguments
lead to the famous Singularity Theorems of GR [2].

In recent times, one has seen the emergence of the string or
membrane viewpoint as the alternative for the point particle.
The motivation for this is to arrive at a viable theory of
quantum gravity  as well as a unification of all forces
. At Planck length scales the stringy nature of the point
particle is believed to remove the problem of renormalizability
of gravity and solve the singularity problem of
GR .

If one accepts the string or membrane viewpoint then one should
be able to write down the corresponding generalized Raychaudhuri
equations for timelike/null worldsheet congruences and arrive at
similar focussing and singularity theorems in Classical String
theory. Very recently, Capovilla and Guven [4] have written down the
generalized Raychaudhuri equations for timelike worldsheet
congruences. In this paper, we construct explicit examples of these
rather complicated set of equations by specializing to certain
simple extremal families of surfaces. Our principal
aim is to extract some information regarding {\em focussing of families
of surfaces} in a way similar to the results for geodesic congruences
in GR.

Sec. II of the paper contains a brief review of the generalized
Raychaudhuri equations {\em a la} Capovilla and Guven.

Sec. III contains the case of extremal string worldsheets in flat and curved
backgrounds. After a general
treatment of the generalized Raychaudhuri equations in string theory
we move on to specific cases. Using the well--known string configurations
in  Rindler spacetime we analyse the resulting Raychaudhuri equation.
For De Sitter spacetime we are able to make some general comments primarily
because of certain specific properties of the spacetime itself.

In the fourth section of the paper we shall deal with the case of
hypersurfaces(i.e. $N-1$ dimensional surfaces embedded in a $N$
dimensional background). We begin by writing down the full set of
generalized Raychaudhuri equations for these objects and then move on to a
discussion of the special case of an
 extremal timelike hypersurfaces (two dimensional) in a curved
Lorentzian background (a wormhole metric in $2+1$ dimensions).
The final section of the paper contains a summary and remarks on future
directions.

 In the Appendix to the paper we present the pedagogical example
of a catenoidal membrane embedded in a three dimensional Euclidean
background. This example, it is hoped, will serve as an useful exercise
while learning about these
equations. It may also turn out to be a relevant calculation
 in the very active area of
biological(amphiphilic) membranes.

\section{FORMALISM}

This section reviews the recent work of Capovilla and Guven[4] which deals with
a generalisation of the
Raychaudhuri Equations for $D$ dimensional surfaces
embedded in an $N$ dimensional background.

We define a $D$ dimensional surface in an $N$ dimensional background through
the embedding $x^{\mu} = X^{\mu}(\xi^{a})$ where $\xi^{a}$ are the coordinates
on the surface and $x^{\mu}$ are the ones in the background. Furthermore, with
the help of an
orthonormal basis $(E^{\mu}_{a}, n^{\mu}_{i})$ consisting of $D$ tangents and
$N-D$
normals we can write down the Gauss--Weingarten equations using the
usual definitions of extrinsic curvature, twist potential and the
worldsheet Ricci rotation coefficients.

In order to analyse deformations normal to the worldsheet we need to
consider the normal gradients of the spacetime basis set.
The corresponding analogs of the Gauss--Weingarten
equations are :

\begin{eqnarray}
D_{i}E_{a} = J_{aij}n^{j} + S_{abi}E^{b} \\
D_{i}n_{j} = -J_{aij}E^{a} + {\gamma}^{k}_{ij}n_{k}
\end{eqnarray}

where $D_{i} \equiv n^{\mu}_{i} D_{\mu}$ ( $D_{\mu}$ being the usual
spacetime covariant derivative). The quantities $J_{a}^{ij}$, $S_{abi}$ and
 ${\gamma}^{k}_{ij}$ are defined as :

\begin{eqnarray}
S^{i}_{ab} = {g}_{\mu\nu}n^{\alpha
i}(D_{\alpha}E^{\mu}_{a})E^{\nu}_{b} \\
{\gamma}^{k}_{ij} = {g}_{\mu\nu}n^{\alpha}_{i}
(D_{\alpha}n^{\mu}_{j})n^{\nu k} \\
J_{a}^{ij} = {g}_{\mu\nu}n^{\alpha i}(D_{\alpha}E^{\mu}_{a})n^{j\nu}
\end{eqnarray}

The full set of equations governing the evolution of deformations
can now be obtained by taking the worldsheet gradient of $J_{a}^{ij}$

\begin{equation}
{\tilde{\nabla}}_{b}J^{ij}_{a} = -{\tilde{\nabla}}^{i}K_{ab}^{j}
-{J_{b}^{i}}_{k}{J_{a}^{kj}} - K_{bc}^{i}K^{cj}_{a} - g(R(E_{b},n^{i})E_{a},
n^{j})
\end{equation}

where the extrinsic curvature tensor components are $K_{ab}^{i} =
-g_{\mu\nu}E^{\alpha}_{a}(D_{\alpha}E^{\mu}_{b})n^{\nu i}$

On tracing over worldsheet indices we get
\begin{equation}
{\tilde{\nabla}}_{a}J^{aij} = -{J_{a}^{i}}_{k}{J^{akj}} - K_{ac}^{i}K^{acj}
 - g(R(E_{a},n^{i})E^{a},
n^{j})
\end{equation}

where we have used the equation for extremal membranes (i.e. $K^{i} = 0$)

The antisymmetric part of (6) is given as :

\begin{equation}
{\tilde{\nabla}}_{b}J^{ij}_{a}- {\tilde{\nabla}}_{a}J^{ij}_{b}
= G_{ab}^{ij}
\end{equation}

where $g(R(Y_{1},Y_{2})Y_{3},Y_{4}) = R_{\alpha\beta\mu\nu}Y_{1}^{\alpha}
Y_{2}^{\beta}Y_{3}^{\mu}Y_{4}^{\nu}$ and

\begin{equation}
G_{ab}^{ij} = -{J_{b}^{i}}_{k}{J_{a}^{kj}} - K_{bc}^{i}K^{cj}_{a} -
g(R(E_{b},n^{i})E_{a},
n^{j}) - (a\rightarrow b)
\end{equation}

One can further split $J_{aij}$ into its symmetric traceless, trace and
antisymmetric parts ($J_{a}^{ij} = {\Sigma}_{a}^{ij} + {\Lambda}_{a}^{ij}
+ \frac{1}{N-D}{\delta}^{ij}{\theta_{a}}$) and obtain the evolution equations
for each of these quantities. The one we shall be concerned with mostly
is given as

\begin{equation}
\Delta \gamma + {\frac{1}{2}}{\partial}_{a}\gamma{\partial}^{a}\gamma +
(M^{2})^{i}_{i} = 0
\end{equation}

with the quantity $(M^{2})^{ij}$ is given as :

\begin{equation}
(M^{2})^{ij} = K_{ab}^{i}K^{abj} +
R_{\mu\nu\rho\sigma}E^{\mu}_{a}n^{\nu i}E^{\rho a}n^{\sigma j}
\end{equation}

 ${\nabla}_{a}$ is the worldsheet covariant derivative ($\Delta = {\nabla^{a}
 \nabla_{a}}$) and ${\partial}_{a}\gamma={\theta}_{a}$.
  Notice that
 we have set ${\Sigma}_{a}^{ij}$ and
${\Lambda}^{ij}_{a}$ equal to zero. This is possible only if the symmetric
traceless part of $(M^{2})^{ij}$ is zero. One can check this by looking
at the full set of generalized Raychaudhuri equations involving
$\Sigma^{ij}_{a}$, ${\Lambda}^{ij}_{a}$ and $\theta_{a}$ [4].
For geodesic curves the usual Raychaudhuri
equations can be obtained by noting that $K^{i}_{00} = 0$, the $J_{aij}$ are
related to their spacetime counterparts $J_{\mu\nu a}$ through
the equation $J_{\mu\nu a} = n^{i}_{\mu}n^{j}_{\nu}J_{aij}$,
and the $\theta$ is defined by contracting with the projection
tensor $h_{\mu\nu}$.

The $\theta _{a}$ or $\gamma$ basically tell us how the
spacetime basis vectors change along the normal directions  as we
move along the surface. If ${\theta}_{a}$ diverges somewhere
, it induces a divergence in $J_{aij}$ , which, in turn means
that the gradients of the spacetime basis along the normals have
a discontinuity. Thus the family of worldsheets meet along a
curve and a cusp/kink is formed. This, we claim, is a
 focussing effect for extremal surfaces analogous to
 geodesic focussing in GR where families of geodesics focus at apoint
 if certain specific conditions on the matter stress energy are
 obeyed.

We now move on to the discussion of the special cases.

\section{STRING WORLDSHEETS}

Two dimensional timelike surfaces embedded in a four dimensional
background are the objects of discussion in this section. We begin
by writing down the generalised Raychaudhuri equation  for the case
in which $\Sigma_{a}^{ij}$ and $\Lambda_{a}^{ij}$ are set to zero
(i.e implicitly assuming that $(M^{2})^{ij}$ does not have a nonzero
symmetric traceless part.Thus we have

\begin{equation}
- \frac{{\partial}^{2}F}{\partial{\tau}^{2}} +
\frac{{\partial}^{2}F}{\partial{\sigma}^{2}}
+\Omega^{2}(\sigma,\tau)  (M^{2})^{i}_{i}(\sigma,\tau)F = 0
\end{equation}

where ${\Omega}^{2}$ is the conformal factor of the induced
metric written in isothermal coordinates.
Notice that the above equation is a second--order, linear, hyperbolic
partial differential equation. On the contrary, the Raychaudhuri equation for
curves is an linear, second order, ordinary differential equation.
The easiest way to analyse the solutions of this
equation with respect to focussing is to assume separability of the
quantity $\Omega^{2}(M^{2})^{i}_{i}$. Then, we have

\begin{eqnarray}
\Omega^{2}(M^{2})^{i}_{i} = M^{2}_{1}(\tau) + M^{2}_{2}(\sigma)\\
F(\tau,\sigma) = F_{1}(\tau) \times F_{2}(\sigma)
\end{eqnarray}

With these we can now split the partial differential equation
into two ordinary differential equations given by

\begin{eqnarray}
\frac{d^{2}F_{1}}{d\tau^{2}} + (\omega^{2} - M^{2}_{1}(\tau))F_{1} = 0 \\
\frac{d^{2}F_{2}}{d\sigma^{2}} + (\omega^{2} + M^{2}_{2}(\sigma))F_{2} = 0
\end{eqnarray}

Since the expansions along the $\tau$ and $\sigma$ directions can be
written as $\theta_{\tau} = \frac{\dot F_{1}}{F_{1}}$ and $\theta_{\sigma}
=\frac{F_{2}^{\prime}}{F_{2}}$ we can analyse focussing effects by
locating the zeros of $F_{1}$ and $F_{2}$ much in the same way as for
geodesic curves [5].
The well--known theorems on the existence of zeros
of ordinary differential equations as discussed in [5]
make our job much simpler. The theorems essentially state that the solutions
of equations of the type $\frac{d^{2}A}{dx^{2}} + H(x)A = 0$ will have
at least one zero iff $H(x)$ is positive definite.
Thus for our case here, we can conclude that, focussing
along the $\tau$ and $\sigma$ directions will take place only if

\begin{equation}
\omega^{2} \ge max[M^{2}_{1}(\tau)]
\qquad ; \qquad \omega^{2} \ge max[-M^{2}_{2}(\sigma)]
\end{equation}

 For stationary strings, one notes that $(M^{2})^{ij}$ will
not have any dependence on $\tau$. Thus we can set $M_{1}^{2}$ equal to
zero. Thus, focussing will entirely depend on the sign of the quantity
$M^{2}_{2}$. We can write $M^{2}_{2}$ alternatively as follows. Consider the
Gauss--Codazzi integrability condition:

\begin{equation}
R_{\mu\nu\alpha\beta}E^{\mu}_{a}E^{\nu}_{b}E^{\alpha}_{c}E^{\beta}_{d}
= R_{abcd} - K_{aci}K^{aci} + K_{bdi}K^{bdi}
\end{equation}

Trace the above expression on both sides with $\eta^{ac}\eta^{bd}$ and
reaarange terms to obtain :

\begin{equation}
K_{abi}K^{abi} = -^{2}R + K^{i}K_{i}
+ R_{\mu\nu\alpha\beta}E^{\mu}_{a}E^{\alpha a}
E^{\nu}_{b}E^{\beta b}
\end{equation}

Thereafter, use this expression and the fact that $n^{i\mu}n^{\nu}_{i}
= g^{\mu\nu} - E^{\mu a}E^{\nu}_{a}$ in the original expression for
$(M^{2})^{i}_{i}$ (see Eqn.(11)) to get (for extremal configurations
with $K^{i}= 0$)

\begin{equation}
M^{2}_{2} = - {}^{2}R + R_{\mu\nu}E^{\mu a}E^{\nu}_{a}
\end{equation}

One can notice the following features from the above expression:

(i) If the background spacetime is a vacuum solution of the Einstein
equations then the positivity of  $M^{2}_{2}$ is guaranteed iff
$^{2}R \le  0$. Thus all string configurations in vacuum spacetimes
which have negative Ricci curvature everywhere will necessarily imply
focussing. This includes the well known string solutions in
Schwarzschild and Kerr backgrounds.

(ii) If the background spacetime is a solution of the Einstein equations
then we can replace the second and third terms in the expressions
for $M^{2}_{2}$ by the corresponding ones involving the Energy
Momentum tensor $T_{\mu\nu}$ and its trace. Thus we have

\begin{equation}
M^{2}_{2} =\left( - \frac{1}{2}g_{\mu\nu}{}^{2}R + T_{\mu\nu} - \frac{1}{2}
Tg_{\mu\nu}\right )E^{\mu a}E^{\nu}_{a}
\end{equation}

Notice that if we split the quantity $E_{a}^{\mu}E^{\nu a}$ into two terms
such as $E_{\tau}^{\mu}E^{\nu \tau}$ and $E_{\sigma}^{\mu}E^{\nu \sigma}$
then we have :

\begin{equation}
M^{2}_{2} = - ^{2}R + \left (T_{\mu\nu} - \frac{1}{2}
Tg_{\mu\nu}\right )E^{\mu \tau}E^{\nu}_{\tau} + \left (T_{\mu\nu} - \frac{1}{2}
Tg_{\mu\nu}\right )E^{\mu \sigma}E^{\nu}_{\sigma}
\end{equation}
 The second term in the above equation is the L. H. S. of the
 Strong Energy Condition (SEC).
Apart from this we have two other terms which are entirely dependent on the
fact that we are dealing with extended objects. The positivity of the
whole quantity can therefore be thought of as an {\em Energy Condition}
for the case of strings. Thus even if the background spacetime
satisfies the SEC, focussing of string world--sheets is not
guaranteed--worldsheet curvature and the projection of the combination
$T_{\mu\nu} - \frac{1}{2}g_{\mu\nu}T$ along the $\sigma$
direction have an important role to play in deciding  focussing/defocussing.

Let us now try to understand the consequences of the above equations for
certain specific flat and curved backgrounds for which the string solutions
are known.

\subsection{Rindler Spacetime}

The metric for four dimensional Rindler spacetime is given as

\begin{equation}
ds^{2} = -a^{2}x^{2}dt^{2} + dx^{2} + dy^{2} + dz^{2}
\end{equation}

We recall from [6] the a string solution in a Rindler spacetime:

\begin{eqnarray}
t=\tau \quad ; \quad x = ba\cosh {a\sigma_{c}} \quad ;{\nonumber} \\ \quad
y = ba^{2}\sigma_{c} \quad ; \quad z = z_{0} \quad (constant)
\end{eqnarray}

where $d\sigma_{c}=\frac{d\sigma}{{a^{2}x^{2}}}$ and $b$ is an integration
constant.
The orthonormal set of tangents and normals to the worldsheet  can be chosen
to be as  follows:

\begin{equation}
E_{\tau}^{\mu} \equiv \left ( \frac{1}{ax}, 0, 0, 0 \right ) \quad ; \quad
E_{\sigma}^{\mu} \equiv \left ( 0, \tanh {a\sigma_{c}}, sech {a\sigma_{c}}, 0
\right )
\end{equation}

\begin{equation}
n^{\mu}_{1} \equiv \left ( 0, 0, 0, 1 \right ) \quad ; \quad
n^{\nu}_{2} \equiv \left ( 0, sech a\sigma_{c}, - \tanh a\sigma_{c}, 0 \right )
\end{equation}

In the worldsheet coordinates $\tau , \sigma_{c}$ the induced metric is flat
and the components of the extrinsic curvature tensor turn out to be

\begin{eqnarray}
K^{1}_{ab} = 0 \quad ; \quad K^{2}_{\tau\tau} = -K^{2}_{\sigma_{c}\sigma_{c}}
 = \frac{1}{ba \cosh^{2}a\sigma_{c}} \quad ; {\nonumber} \\
   K^{2}_{\sigma\tau} = 0
\end{eqnarray}

The quantity $(M^{2})^{i}_{i}$ which is dependent only on the extrinsic
curvature
of the worldsheet (the background spacetime being flat) turns out to be

\begin{equation}
(M^{2})_{i}^{i} = \frac{2}{b^{2}a^{2}\cosh^{4}a\sigma_{c}}
\end{equation}

Therefore the generalized Raychaudhuri equation turns out to be

\begin{equation}
-\frac{{\partial}^{2}F}{{\partial}\tau^{2}} +
\frac{{\partial}^{2}F}{{\partial}\sigma_{c}^{2}} + \frac{2a^{2}}{\cosh^{2}
a\sigma_{c}} F
= 0
\end{equation}

Separating variables ($F= T(\tau)\Sigma(\sigma)$ ) one gets the harmonic
oscillator
equation for $T$ and the Poschl Teller equation for positive eigenvalues [7]
for $\Sigma$ which is given as:

\begin{equation}
\frac{d^{2}\Sigma}{d\sigma^{2}} + \left ( {\omega}^{2} +
\frac{2a^{2}}{\cosh^{2}\sigma} \right ) \Sigma = 0
\end{equation}

{}From the results of Tipler [5] on the zeros of differential equations
one can conclude that focussing will occur ($H(\sigma) > 0$ always)

\subsection{De Sitter Spacetime}

The metric for De Sitter spacetime is given as :

\begin{equation}
ds^{2} = - f(r)dt^{2} + \frac{dr^{2}}{f(r)} + r^{2}\left ({d\theta^{2} +
\sin^{2}
\theta d\phi^{2}}\right )
\end{equation}

with $f(r) = 1- H^{2}r^{2}$.

Since De Sitter spacetime is an Einstein space we have a clear advantage.
The Riemann, Ricci tensors can be written as

\begin{equation}
R_{\mu\nu\alpha\beta} = H^{2} \left ( g_{\mu\alpha}g_{\nu\beta} - g_{\mu\beta}
g_{\nu\alpha} \right ) \qquad ; \quad R_{\mu\nu} = 3H^{2}g_{\mu\nu}
\end{equation}

Therefore a little bit of calculation will reveal that the Raychaudhuri
equation for all string configurations in DeSitter space can be written
as :

\begin{equation}
-\frac{{\partial}^{2}F}{{\partial}\tau^{2}} +
\frac{{\partial}^{2}F}{{\partial}\sigma_{c}^{2}}
+ \Omega^{2} \left ( -{}^{2}R + 6H^{2} \right ) F = 0
\end{equation}

Focussing in this case thus depends only on the Ricci scalar and the
conformal factor $\Omega^{2}$ of the
worldsheet metric. One can write down the solutions of the above equation
for the various string configurations in De Sitter space mentioned in [6].
This is simple enough and we shall refrain from writing them out explicitly.

Some other examples have been constructed in [8]. These involve the case of a
$3+1$ dimensional Lorentzian wormhole geometry as the background.

\section{HYPERSURFACES}

We now move on to the special case of timelike hypersurfaces.
Here we have $D$ quantities $J_{a}$ but only one normal
defined at each point on the surface. The Eqn. (8) turns out to
be :

\begin{equation}
{\partial}_{b}J_{a} - {\partial}_{a}J_{b} = 0
\end{equation}

Therefore one can write $J_{a} = {\partial}_{a}\gamma$ and the traced
equation (7) becomes,

\begin{equation}
\Delta \gamma + ({\partial}_{a}\gamma )( {\partial}^{a}{\gamma}) + M^{2} = 0
\end{equation}

with

\begin{eqnarray}
M^{2} =  K_{ab}K^{ab} + R_{\nu\sigma}n^{\nu}n^{\sigma} \nonumber \\
=-^{2}R + R_{\mu\nu}E^{\mu a}E^{\nu}_{a}
=-^{2}R + ^{3}R - R_{\mu\nu}n^{\mu}n^{\nu}
\end{eqnarray}

where we have used $n^{\mu}n^{\nu} = g^{\mu\nu} - E^{\mu}_{a}E^{\nu a}$
and the Gauss--Codazzi integrability condition.

If we assume that the background spacetime satisfies the Einstein equation
then we have:

\begin{equation}
M^{2} = -\left ( ^{2}Rg_{\mu\nu} + T_{\mu\nu} + Tg_{\mu\nu}\right )
n^{\mu}n^{\nu}
\end{equation}

Thus, for stationary two dimensional hypersurfaces (strings in 3D backgrounds)
we have the same conclusions as obtained in the previous section.
For a two--dimensional hypersurface in  three--dimensional
flat background the task is even simpler. $M^{2}$ can be shown to be equal to
the negative of the Ricci scalar of the membrane's induced metric and
$^{2}R\leq 0$ guarantees focussing. We will discuss in detail the case of a 2D
hypersurface(catenoid) embedded in a 3D flat, Euclidean background
in the Appendix to this paper.

Let us now turn to a specific case where the equations are exactly solvable.

\subsection{Hypersurfaces in a $2+1$ Curved Background}

 Our backgound spacetime here is curved,
Lorentzian background and $2+1$ dimensional. The metric we choose is that
of a Lorentzian wormhole in $2+1$ dimensions given as :

\begin{equation}
ds^{2} = -dt^{2} + dl^{2} + \left ( b_{0}^{2} + l^{2}\right ) d\theta^{2}
\end{equation}

A string configuration in this background can be easily found by solving the
geodesic equations in the $2D$ spacelike hypersurface [8]. This turns out to be

\begin{equation}
t=\tau \quad ; \quad l=\sigma \quad ; \quad \theta = \theta_{0}
\end{equation}

The tangents and normal vectors are simple enough:

\begin{equation}
E^{\mu}_{\tau} \equiv \left ( 1, 0, 0  \right )  \quad ; \quad
E^{\mu}_{\sigma} \equiv \left ( 0, 1, 0 \right ) \quad ; \quad
n^{\mu} = \left ( 0, 0, \frac{1}{b^{2}_{0} + l^{2}} \right )
\end{equation}

The extrinsic curvature tensor components are all zero as the induced metric
is flat. Using the Riemann tensor components (which can be evaluated simply
using the standard formula) we can write down the generalised Raychaudhuri
equation. This turns out to be :

\begin{equation}
-\frac{{\partial}^{2}F}{{\partial}\tau^{2}} +
\frac{{\partial}^{2}F}{{\partial}\sigma^{2}}
+ \left (-\frac{b_{0}^{2}}{(b_{0}^{2} + \sigma^{2})^{2}} \right ) F = 0
\end{equation}

A separation of variables $F= T(\tau)\Sigma(\sigma)$ will result in two
equations--one of which is the usual Harmonic Oscillator and the other
given by:

\begin{equation}
\frac{d^{2}\Sigma}{d\sigma^{2}} + \left ( \omega^{2} - \frac{b_{0}^{2}}{(b_{0}
^{2} + \sigma^{2})^{2}} \right ) \Sigma = 0
\end{equation}

The above equation can be recast into the one for Radial Oblate Spheroidal
Functions
by a simple change of variables -- $ \Sigma^{\prime} = \sqrt{b_{0}^{2} +
\sigma^{2}} \Sigma $.

\begin{equation}
(1+ \xi^{2} ) \frac{d^{2}{\Sigma^{\prime}}}{d\xi^{2}} +
 2\xi\frac{d\Sigma^{\prime}}{d\xi} + \left (\omega^{2}b_{0}^{2}(1+\xi^{2})
\right )
 \Sigma^{\prime} = 0
 \end{equation}

where $\xi =\frac{\sigma}{b_{0}}$.

 The general equation for Radial Oblate Spheroidal Functions is given as :

 \begin{equation}
(1+ \xi^{2} ) \frac{d^{2}{V_{mn}}}{d\xi^{2}} +
 2\xi\frac{dV_{mn}}{d\xi} + \left (-\lambda_{mn} +
 k^{2}{\xi}^{2}
 - \frac{m^{2}}{1+\xi^{2}} \right ) V_{mn} = 0
 \end{equation}

Assuming $m=0$ and $ \lambda_{0n} = -k^{2} = -\omega^{2}b_{0}^{2} $
we get the equation for our case. The solutions are finite at infinity
and behave like simple sine/cosine waves in the variable $\sigma$.
Consulting the tables in [9] we conclude that only for $n=0,1$ we
can have $\lambda_{0n}$ to be negative. In general, the scattering
problem for the Schroedinger--like equation has been analysed
numerically in [10].

As regards focussing, one can say from the differential equations and the
theorems stated in [5] that the function $\Sigma^{\prime}$ will always have
zeros if $\omega^{2} \ge \frac{1}{b_{0}^{2}}$. Even from the series
representations (see [9]) of the Radial Oblate Spheroidal Functions we
can exactly locate the zeros and obtain explicitly the focal curves. However,
we shall not attempt such a task here.

\section{CONCLUSIONS}

The basic aim of this paper has been to obtain explicit examples
of the generalised
Raychaudhuri equations derived in [4]. To this end we have discussed two
specific cases--that of string worldsheets and hypersurfaces. In the latter
case
the {\em full} set of equations simplify considerably. We have solved
them for the
case of an extremal $2D$ timelike membrane in a curved $2+1$ dimensional
Lorentzian wormhole background.

In the case of strings we have been able to reduce the generalised
Raychaudhuri equation to a form reminiscent of the one for geodesic
congruences by assuming that the background spacetime satisfies the
Einstein equations with a specific form of matter. Any assumption of an
Energy Condition for the matter generating the background spacetime
does not seem to
lead to focussing effects. The presence of the worldsheet Ricci scalar
and the spacetime Ricci tensor
together control focussing effects. For backgrounds obeying $R_{\mu\nu}=0$
it is necessary to have worldsheets of negative worldsheet curvature in order
to ensure focussing for stationary strings. If the backgrounds obey
the Einstein field equations with a specific energy--momentum tensor
then we have a specific condition which ensures focussing.

However, it is not completely clear what role the extrinsic curvature term
in $(M^{2})^{i}_{i}$ term plays as regards worldsheet focussing in the
general case (i.e. with $\Omega^{2}(M^{2})^{i}_{i}$ not separable).
This requires a more extensive analysis of the general
features of the partial differential equation. Moreover,
if one wishes to conclude about the existence/nonexistence of spacetime
singularities one has to frame the notions of worldsheet completeness
in analogy with geodesic completeness and relate the idea of worldsheet
incompleteness with the presence of a singularity in the background
spacetime.

This, indeed is a fairly difficult problem (infact it does'nt even have a
clear formulation). However, a solution of such a problem would actually
tell us
whether a string description as an alternative to the point particle can
actually lead to the presence/absence of spacetime singularities at the
classical level. From the
fact that focussing is present in almost all the cases under discussion
here we cannot conclusively say anything about spacetime singularities.
The only progress we have been able to make in this paper is to
introduce the idea and the conditions under which focussing can take
place for families of string worldsheets or hypersurfaces.
\vspace{.3in}

{\bf ACKNOWLEDGEMENTS}
\vspace{.2in}

Financial support from the Institute of Physics, Bhubaneswar
in terms of a fellowship is gratefully acknowledged.



\appendix
\section{A PEDAGOGICAL EXAMPLE -- THE CATENOIDAL MEMBRANE}

This appendix contains a pedagogical example of the generalised
Raychaudhuri equations which may help the reader in understanding
the equations better. Moreover, there do exist a whole class of
physically relevant systems which can be modelled using the
theory of 2D surfaces embedded in a Euclidean background [11]. It is hoped
that the example below may serve to be useful in that context also.

The background metric here is flat and Euclidean. It is given as:

\begin{equation}
ds^{2}= dx^{2}+dy^{2}+dz^{2}
\end{equation}

The embedding of a two dimensional surface in this three
dimensional background is specified by three functions
$x(u,v)$, $y(u,v)$ and $z(u,v)$ where $u$ and $v$ are the
coordinates on the surface (worldsheet).

For the catenoid we have:

\begin{equation}
x(u,v) = b_{0}\sinh u \cos v \quad ; \quad y(u,v) = b_{0}\sinh u
\sin v \quad ; \quad z(u,v) = b_{0}v
\end{equation}

The induced metric on the surface ${\gamma}_{ab}$ has only
nonzero diagonal elements:

\begin{equation}
{\gamma}_{uu} = {\gamma}_{vv} = b_{0}^{2}\cosh ^{2}u \quad ;
\quad {\gamma}_{uv} = {\gamma}_{vu} = 0
\end{equation}

We now choose an orthonormal basis for which
$g_{\mu\nu}E_{a}^{\mu}E_{b}^{\nu} = {\delta}_{ab}$
which is convenient for calculating the extrinsic curvature
The tangent vectors and the normal are given as
:

\begin{eqnarray}
E_{u}^{\mu} \equiv (\cos v, \sin v, 0) \quad  {\nonumber}\\
\equiv \quad
(\frac{x}{r}, \frac{y}{r}, 0)
\end{eqnarray}
\begin{eqnarray}
E_{v}^{\mu} \equiv (- \tanh u \sin v, \tanh u \cos v , sech u)
\quad {\nonumber} \\ \equiv \quad (-\frac{y}{r_{1}}, \frac{x}{r_{1}},
\frac{b_{0}}{r_{1}})
\end{eqnarray}
\begin{eqnarray}
n^{\mu} \equiv (\frac{\sin v}{\cosh u}, \frac{\cos v}{\cosh
u},-\tanh u) {\nonumber} \\ \quad \equiv \quad (-\frac{yb_{0}}{rr_{1}},
\frac{xb_{0}}{rr_{1}}, \frac{r}{r_{1}})
\end{eqnarray}

where $r=\sqrt{x^{2}+y^{2}}$ and
$r_{1}=\sqrt{x^{2}+y^{2}+b_{0}^{2}}$.

The extrinsic curvature tensor components are defined as $K_{ab} =
- g_{\mu\nu}E^{\alpha}_{a}(D_{\alpha}E^{\mu}_{b})n^{\nu}$. For the
embedding under consideration here one gets:

\begin{equation}
K_{uu} = K_{vv} = 0 \qquad ; \qquad K_{uv} = K_{vu} =
-\frac{1}{b_{0}\cosh^{2} u}
\end{equation}

With the above expressions we can now straightaway write down
the Raychaudhuri equations. In this case we have only one of these.
 Using $\gamma = \ln F$ and

\begin{equation}
\Delta \equiv \frac{1}{b_{0}^{2}\cosh^{2}u} \left
(\frac{{\partial}^{2}}{{\partial}u^{2}} +
\frac{{\partial}^{2}}{{\partial} v^{2}} \right )
\end{equation}

we get:

\begin{equation}
\frac{{\partial}^{2}F}{{\partial}u^{2}} +
\frac{{\partial}^{2}F}{{\partial}v^{2}} + \frac{2}{\cosh^{2}u} F
= 0
\end{equation}

We can now separate variables using $F(u,v) = U(u)V(v)$ and
obtain two independent equations for $U(u)$ and $V(v)$. These
are :

\begin{equation}
\frac{d^{2}V}{dv^{2}} + {\omega}^{2}V = 0
\end{equation}

\begin{equation}
\frac{d^{2}U}{du^{2}} + \left ( {-\omega}^{2} +
\frac{2}{\cosh^{2}u} \right ) U = 0
\end{equation}

The second of these is the well--known Poschl--Teller equation
which occurs in quantum mechanics [7]. We now write down the
solutions to these equations and obtain the expressions for the
expansions ${J}_{u}$ and ${J}_{v}$. These expansions
will tell us about the nature of the deformations of the
catenoidal membrane.

The solutions to the equation for $V(v)$ (for $\omega \neq 0$) are simple:

\begin{equation}
V(v) = \sin {\omega v} \qquad or \qquad \cos {\omega v}
\end{equation}

Now since $v$ is an angle coordinate we must have $V(v + 2\pi) =
V(v)$ which implies ${\omega}_{n} = {n}$ where $n$ is
integral ($n \neq 0$). For $n=0$ one can easily check that $V(v) = constant$
is the only possible solution (this is once again because $v$ is an angle
coordinate).
  We shall now have to use this input in the equation
for $U$ in order to obtain the relevant solutions.

Before we do that let us evaluate ${J}_{v}$. From the
relation $F= UV $ and $\gamma = \ln F$ we get $\gamma = \ln U +
\ln V$. Thus we have (from $J_{a} =
{\partial}_{a}\gamma $):

\begin{equation}
{J}_{u} = \frac{U^{\prime}}{U} \qquad ; \qquad {J}_{v}
= \frac{V^{\prime}}{V}
\end{equation}

Thus depending on which of the solutions $V(v)$ we choose, we
get a separate  expression for $J_{v}$.

\begin{eqnarray}
V(v) = \sin nv \qquad : \qquad {J}_{v} = n\cot nv \\
V(v) = \cos nv \qquad : \qquad {J}_{v} = -n\tan nv \\
\end{eqnarray}

Thus focussing in the angular direction can occur at $v = \pi
, 2\pi (0)$ ($\sin$ solution) or $v= \frac{\pi}{2} ,
\frac{3\pi}{2}$ ($\cos$ solution). At these points the families of
$u =constant$ closed curves meet.

Now we move over to the solutions of the $U$ equation which are
of course much more nontrivial.
First let us rewrite the $U$ equation in a different form by
introducing the variable $J_{u} = \frac{U^{\prime}}{U}$.
This results in a first order equations of the Riccati type
and is useful in discussing focussing.

\begin{equation}
\frac{d{J}_{u}}{du} + {J}_{u}^{2} = -\left (
-{n}^{2} + \frac{2}{\cosh^{2}u} \right )
\end{equation}

This equation is similar to the Raychaudhuri equation in GR.
Thus, if the R.H.S is negative then we can get focussing.
For $n = 0 $ there seems to be no problem with focussing
whereas for $ n = 1$ focussing is possible only within a
finite region of $u$ (as the R.H.S. is negative only in that
domain. For all $n\ge 2$ one does not get any focussing
(the R.H.S is positive for all $u$).

The solutions to the Poschl--Teller equations can be obtained in
terms of Hypergeometric Functions [7]. Since, in this case
we are interested in the Poschl--Teller equation only as a
differential equation and not as a potential problem in
quantum mechanics we shall be concerned with solutions which are
finite everywhere as well as those which diverge at specific
values of the independent variable.

We now list the various solutions for different $n$ and the
corresponding $J_{u}$.

{\bf n = 0}\quad : \quad
{\em First Solution}

\begin{equation}
U_{1}(u) = F[-1,2,1,\frac{1}{2}(1-\tanh u)] = P_{1}(\tanh u)
    = \tanh u
\end{equation}
\begin{equation}
{J}_{u}^{(1)} = \frac{1}{\sinh u \cosh u}
\end{equation}

{\bf n = 0}\quad : \quad
{\em Second Solution}

\begin{equation}
U_{2}(u) = u\tanh u - 1
\end{equation}
\begin{equation}
{J}_{u}^{(2)} = \frac{2}{\sinh 2u} + \frac{\coth^{2}u}{u -
{\coth u}}
\end{equation}

{\bf n = 1}\quad : \quad
{\em First Solution}

\begin{equation}
U_{1}(u) = sech u F[0,3,2,\tanh u] = sech u
\end{equation}
\begin{equation}
{J}_{u}^{(1)} = -\tanh u
\end{equation}

{\bf n = 1}\quad : \quad {\em Second Solution}

\begin{equation}
U_{2}(u) = \frac{u + \sinh u \cosh u}{\cosh u}
\end{equation}

\begin{equation}
{J}_{u}^{(2)} = -\tanh u + \frac{1+\cosh 2u}{u + \sinh u
\cosh u}
\end{equation}

It is worthwhile to point out that the second solutions have
been obtained in both cases by solving the nonlinear first order
equation involving $J_{u}$ (Riccati equation). We have used
the well known fact that if one solution of a Riccati equation
is known, one can derive a second solution by writing it as a
sum of the known one and an unknown function. (The differential
equation in the unknown function reduces to a Bernoulli equation).
Thus obtaining $J_{u}$ we integrate to get $U(u)$.

What do the $J_{u}$ obtained above imply?
For $n=0$ both the solutions have the property that at $u = 0$
${J}_{u} \rightarrow -\infty$ from below (i.e. $u$ negative)
and $J_{u} \rightarrow \infty$ from above. Thus $u = 0$ is
a focal curve( a circle in this case). Additionally, the second
solution has the
intriguing feature that at two symmetrically placed points
(which are solutions to the transcendental equation $ u =\coth
u$), ${J}_{u}$ has exactly similar behaviour. These are
focal curves too. In the $n =1$ case however one of the
solutions (the first one) does not lead to any focussing at
all-- we get an almost parallel family of surfaces. The other
solution for $n = 1$ indicates focussing only at $u = 0$ and
nowhere else. If one is inclined to consider a membrane of
finite extent (such as a soap film formed between coaxial rings
placed a certain distance apart [12]) then one needs to have
$J_{a}$ diverging at a finite value of $u$. This seems
possible only for the second solution for $n = 0$.

Another extremal two--dimensional surface (embedded in three dimensional
Euclidean space) is the helicoid. Interestingly there exists a local isometry
of the helicoid into the catenoid. If $u_{1}$ and $v_{1}$ are the
coordinates on the helicoid this is given as :

\begin{equation}
v_{1} =v \qquad u_{1} = \sinh u
\end{equation}

Thus, one can essentially use the same pair of differential equations
for $u$ and $v$ given above for the catenoid. However there is one striking
difference. For the helicoid $v_{1}$ is no longer an angle variable
(the ranges of $u_{1}$ and $v_{1}$ are $0<u_{1}<\infty$ and $-\infty
< v_{1} \infty$. Therefore, we do not have a restriction on the allowed
values on $\omega^{2}$ arising solely from the equation for $v_{1}$.
The equation for the $u_{1}$ variable however yields a restriction
$\omega^{2} \leq \frac{3}{2}$ if one is interested in focussing effects.

\end{document}